\documentclass[12pt]{article}
\usepackage[english]{babel}
\usepackage[T1]{fontenc}
\usepackage{amsmath}
\usepackage{amsfonts}

\newlength{\dinwidth}
\newlength{\dinmargin}
\def\RR{{\mathbb R}}

\def\ti{\tilde}

\def\A{{\cal A}}
\def\B{{\mfa}}

\def\H{{\cal H}}

\def\S{{\cal S}}

\def\S2{S^{1(2)}}

\def\<{\langle}
\def\>{\rangle}

\newcommand{\mcJ}{\mathcal J}
\newcommand{\mcI}{\mathcal I}
\newcommand{\ha}{\hat}
\newcommand{\spec}{\te{sp}}
\newcommand{\tas}{\te{as} }
\newcommand{\slim}{\te{s}\textrm{-}\lim}

\newcommand{\ph}{\phantom}
\newcommand{\mcP}{\mathcal P}
\newcommand{\mcR}{\mathcal R}
\newcommand{\pa}{\partial}
\newcommand{\mcD}{\mathcal{D}}
\newcommand{\Q}{Q_{\ka}}
\newcommand{\ka}{\kappa}
\newcommand{\kpout}{\overset{\tout}{\times}_{\ka}}
\newcommand{\kpin}{\overset{\tin}{\times}_{\ka}}
\newcommand{\pout}{\overset{\tout}{\times}}
\newcommand{\pin}{\overset{\tin}{\times}}
\newcommand{\tin}{\te{in}}
\newcommand{\tout}{\te{ out}}
\newcommand{\mcW}{\mathcal W}

\newcommand{\Om}{\Omega}

\newcommand{\h}{\fr{1}{2}}
\newcommand{\te}{\mathrm}

\newcommand{\nat}{\mathbb{N}}
\newcommand{\hil}{\mathcal{H}}

\newcommand{\mfa}{\mathfrak{A}}

\newcommand{\eps}{\varepsilon}
\newcommand{\fr}[2]{\frac{#1}{#2}}
\newcommand{\al}{\alpha}

\newcommand{\real}{\mathbb{R}}

\newcommand{\la}{\lambda}

\newcommand{\non}{\nonumber}

\def\proof{\noindent{\bf Proof. }}
\def\qed{$\Box$\medskip}
\newtheorem{theoreme}{Theorem } [section]
\newtheorem{proposition}[theoreme]{Proposition}
\newtheorem{lemma}[theoreme]{Lemma}
\newtheorem{definition}[theoreme]{Definition}
\newtheorem{corollary}[theoreme]{Corollary}
\newtheorem{remark}[theoreme]{Remark}
\newtheorem{example}[theoreme]{Example}
\newtheorem{criterion}[theoreme]{Criterion}
\newcommand{\beq}{\begin{equation}}
\newcommand{\eeq}{\end{equation}}
\newcommand{\beqa}{\begin{eqnarray}}
\newcommand{\eeqa}{\end{eqnarray}}
\newcommand{\ben}{\begin{arabicenumerate}}
\newcommand{\een}{\end{arabicenumerate}}
\newcommand{\bex}{\begin{example}}
\newcommand{\eex}{\end{example}}
\newcommand{\ber}{\begin{remark}}
\newcommand{\eer}{\end{remark}}
\newcommand{\bec}{\begin{corollary}}
\newcommand{\eec}{\end{corollary}}
\newcommand{\bed}{\begin{definition}}
\newcommand{\eed}{\end{definition}}
\newcommand{\bep}{\begin{proposition}}
\newcommand{\eep}{\end{proposition}}
\newcommand{\becr}{\begin{criterion}}
\newcommand{\eecr}{\end{criterion}}
\def\bel{\begin{lemma}}
\def\eel{\end{lemma}}
\def\bet{\begin{theoreme}}
\def\eet{\end{theoreme}}
\def\bed{\begin{definition}}
\def\eed{\end{definition}}

\begin{document}
\title{Asymptotic completeness in a class of massless relativistic quantum field theories}

\author{
{\bf Wojciech Dybalski\footnote{Supported by the DFG grant SP181/25-1.}}\\
Zentrum Mathematik, Technische Universit\"at M\"unchen,\\
D-85747 Garching, Germany\\
E-mail: {\tt dybalski@ma.tum.de}
\and
{\bf Yoh Tanimoto\footnote{Supported in part by the ERC Advanced Grant 227458
OACFT ``Operator Algebras and Conformal Field Theory''.}}\\
Dipartimento di Matematica, Universit\`a di Roma ``Tor
Vergata''\\ Via della Ricerca Scientifica, 1 - I--00133 Roma, Italy.\\
E-mail: {\tt tanimoto@mat.uniroma2.it}}
\date{}
\maketitle

\abstract{This paper presents the first examples of massless relativistic quantum field
theories which are interacting and asymptotically complete. These two-dimensional  theories 
are obtained  by an application of a deformation procedure, introduced recently by Grosse and Lechner, to
chiral conformal quantum field theories. The resulting models may not be strictly local,
but they contain observables localized in spacelike wedges. It is shown that the 
scattering theory for waves in two dimensions, due to Buchholz, is still valid under these weaker assumptions.
The concepts of interaction and asymptotic completeness, provided by this theory, are adopted
in the present investigation. }

\section{Introduction}
The interpretation of quantum field theories in terms of particles is a long-standing fundamental 
problem \cite{Ha58, Ru62, LSZ55, CD82}. 
The  last two decades  witnessed significant progress on this issue, both on the side of structural analysis
\cite{Bu90, Po04.1, Po04.2, Dy05, Dy09} and in the study of concrete models \cite{Sp97,DG99, FGS04, Le08}. 
By combining  methods of algebraic quantum field theory \cite{Ha} with insights from the form-factor program \cite{SW00,BFK06},  
the first  examples of local, relativistic quantum field theories, 
which are interacting and asymptotically complete, have been constructed in \cite{Le08}  . 
As this class contains only massive models, the question of asymptotic completeness in the
presence of massless particles is open to date in the local, relativistic framework. This 
can be partly attributed to the infamous \emph{infrared problem}, which hinders rigorous construction and analysis
of interacting massless theories by traditional methods (see however \cite{CRW85, BFM04,ZZ92}). It is therefore remarkable that more
recent constructive tools, developed in \cite{BLS10}, give rise to massless models which are asymptotically complete and 
interacting.  We exhibit such theories in the present work.

We recall that a new class of  relativistic quantum field theories, including both massive and massless models,  
has been  obtained recently by a certain deformation procedure akin to the Rieffel  deformation \cite{GL08,BS08,BLS10,DLM10}.  
These theories are wedge-local i.e. observables can be localized
in (unbounded)  wedge-shaped regions extending in spacelike directions. In the massive case  this 
remnant of locality suffices for a canonical construction of the two-body scattering matrix, as shown
in \cite{BBS01}. Exploiting this fact, it was demonstrated in \cite{GL08,BS08} that the deformed theory
is interacting even if the original theory is not. As in general only two-body scattering states are available,
it may seem that the problem of asymptotic completeness cannot be  addressed in the framework of wedge-local theories. 
However, in the case of two-dimensional massless theories such a conclusion would be pre-mature, as we demonstrate in this  paper.

Our first task is to provide a scattering theory for such models.
We recall that for \emph{local} two-dimensional theories of massless excitations a scattering theory 
was  developed in \cite{Bu75}. The basic building blocks of this construction are the subspaces $\hil_+$
and $\hil_-$ in the physical Hilbert space $\hil$, corresponding to the right and left branch of
the lightcone in momentum space. These subspaces carry representations of the Poincar\'e group
which are in general highly reducible. Thus vectors $\Psi_{\pm}\in \hil_{\pm}$ do not describe
particles in the Wigner sense, but rather composite objects, called in \cite{Bu75} 'waves'. 
In view of their dispersionless motion, a composition of several waves traveling in the
same direction (say elements of  $\hil_+$), gives rise to another wave from $\hil_+$. Thus it suffices
to consider scattering states $\Psi_+\pout\Psi_-$ (resp. $\Psi_+\pin\Psi_-$) which describe
two waves traveling in the opposite directions in the remote future (resp. past). They span the subspaces
$\hil^{\tout}$ (resp. $\hil^{\tin}$) of the outgoing (resp. incoming) states. 
The scattering operator $S:\hil^{\tout}\to \hil^{\tin}$ can be defined as an isometry mapping $\Psi_+\pout\Psi_-$ into $\Psi_+\pin\Psi_-$. 
If $\hil^{\tout}=\hil^{\tin}=\hil$, then we say that the theory is asymptotically complete.
As we will show,  there exists a large class of non-interacting massless theories  in two-dimensional spacetime which have this property: it includes all chiral conformal quantum field theories.

In the light of the above discussion, it is not surprising that the scattering theory  from \cite{Bu75} can be generalized to the wedge-local context. Indeed, observables localized in two opposite spacelike wedges suffice to separate two waves traveling in opposite directions. We demonstrate
this fact in Section~\ref{scattering} after some introductory remarks on wedge-local quantum
field theories. In Section~\ref{deformation} we express the scattering matrix $S_{\ka}$ of the 
deformed theory (with a deformation parameter $\ka$), by the scattering matrix $S$ of the original one. We obtain
\beq
S_{\ka}=e^{i\ka M^2}S,
\eeq
where $M$ is the mass operator. Hence, similarly as in the massive case,  the deformed theory is interacting, even  if the original theory is not. Moreover, the property of asymptotic completeness  is preserved by the deformation procedure. Thus, as we show in Section~\ref{chiral}, deformations of chiral conformal field theories give rise to wedge-local theories which are interacting and asymptotically complete.  We summarize our results in Section~\ref{conclusions}, where also some open questions are discussed.

%%%%%%%%%%%%%%%%%%%%%%%%%%%%%%%%%%%%%%%%%%%%%%%%%%%%%%%%%%%%%%%%%%%%%%%%   
\section{Scattering theory}\label{scattering}
%%%%%%%%%%%%%%%%%%%%%%%%%%%%%%%%%%%%%%%%%%%%%%%%%%%%%%%%%%%%%%%%%%%%%%%

A convenient framework for the study of  wedge-local theories is provided  by the concept of a Borchers triple \cite{Bo92}. We recall that a Borchers triple $(\mcR,U,\Om)$, (relative to the wedge $\mcW=\{\, x=(x^0,x^1)\in \real^2\,|\, x^1\geq |x^0|\,\}$), 
consists of:
%%%%%%%%%%%%%%%%%%%%%%%%%%%%%%%%%%%%%%%%%%%%%%%%%%%%%%%%%%%%%%%%%%%%%
\begin{enumerate}  

\item[(a)] a von Neumann algebra $\mcR\subset B(\hil)$,

\item[(b)] a strongly continuous unitary representation $U$ of $\real^2$ on $\hil$,
whose spectrum $\spec\, U$ is contained in the closed forward lightcone 
$V_+=\{\, p=(p^0,p^1)\in\real^2\,|\, p^0\geq |p^1| \,\}$ and which satisfies $\al_x(\mcR)\subset \mcR$,
for $x\in \mcW$, where $\al_x(\,\cdot\,)=U(x)\,\cdot\,U(x)^{-1}$,

\item[(c)] a unit vector $\Om\in\hil$ which is invariant under the action of $U$ and is
cyclic and separating for $\mcR$. It will be called the vacuum vector.

\end{enumerate}
%%%%%%%%%%%%%%%%%%%%%%%%%%%%%%%%%%%%%%%%%%%%%%%%%%%%%%%%%%%%%%%%%%%%%
One interprets $\mfa(\mcW):=\mcR$ as the algebra of all observables localized
in the wedge $\mcW$. In view of (c), one can apply to $(\mcR,\Om)$ the Tomita-Takesaki theory and we denote by $(\Delta,J)$ the modular operator and the conjugation. As shown in \cite{Bo92}, with the help of the modular
objects one can construct an (anti-)unitary representation $\la\to \ti U(\la)$ of the proper Poincar\'e group $\mcP_+$
which extends the original representation of translations. In particular, $J$ implements the spacetime reflection i.e.
\beq
JU(x)J=U(-x), \quad x\in\real^2. \label{inversion}
\eeq 
Thus with any wedge $\la \mcW$ one can associate the algebra of observables $\mfa(\la\mcW)=\ti U(\la)\mcR \ti U(\la)^{-1}$.  
Since,  by the Tomita-Takesaki theory, $J\mcR J=\mcR'$, the resulting net is
wedge-local i.e $\mfa((\la\mcW)')=\mfa(\la\mcW)'$, where $(\la\mcW)'$ is the causal complement of $\la\mcW$ and a prime over an algebra denotes the commutant. Hence this net  gives rise to a (two-dimensional) wedge-local, relativistic quantum field theory. See \cite{Bo92, Fl98} for proofs of the above statements and \cite{BLS10} for a more detailed overview.

Let $(H,P)$ be the generators of $U$ i.e. $U(x^0,x^1)=e^{iHx^0-iPx^1}$.
We set $\hil_{\pm}=\ker\, (H\mp P)$ and denote by $P_{\pm}$ the corresponding projections.
We assume that  $\hil_+\cap\hil_-=[c\Om]$ i.e. $\Om$ is the  unique (up to a phase) unit vector   
which is invariant under translations. This
implies that $\hil_+\cap[c\Om]^\bot$ is orthogonal to $\hil_-\cap[c\Om]^\bot$.
We assume that the latter two subspaces are non-trivial, to ensure that the theory contains massless excitations. 
Let us now describe briefly their collision theory. The construction follows closely  \cite{Bu75}.

For any $F\in B(\hil)$ and $x\in\real^2$ we denote $F(x):=\al_x(F)$ and define the sequences of operators
\beq
F_{\pm}(h_T)=\int dt\,h_T(t)F(t_{\pm})\textrm{ with } t_{\pm}=(t,\pm t),
\eeq
where $h_T(t)=|T|^{-\eps}h(|T|^{-\eps}(t-T))$, $0<\eps<1$ and $h\in C_0^{\infty}(\real)$ is a non-negative, symmetric  function
s.t. $\int dt\, h(t)=1$. With the help of these approximating sequences we construct the asymptotic fields corresponding to the wedge~$\mcW$.
%%%%%%%%%%%%%%%%%%%%%%%%%%%%%%%%%%%%%%%%%%%%%%%%%%%%%%%%%%%%%%%%%%%%%%%%%%%%%%%%%%%%%%%%%
\bel\label{right-wedge-lemma} Let $F\in \mcR$.  Then the limits
\beqa
\Phi_+^{\tout}(F):=\underset{T\to\infty}{\slim}\textrm{ }F_{+}(h_T), \quad
\Phi_-^{\tin}(F):=\underset{T\to-\infty}{\slim}\textrm{ }F_{-}(h_T),\label{minus-in}
\eeqa
exist and are elements of $\mcR$. They depend only on the respective vectors $\Phi_+^{\tout}(F)\Om=P_+F\Om$, $\Phi_-^{\tin}(F)\Om=P_-F\Om$ and  satisfy
\begin{enumerate}
\item[(a)] $\Phi_+^{\tout}(F)\hil_+\subset\hil_+, \quad\Phi_-^{\tin}(F)\hil_-\subset\hil_-$,

\item[(b)] $\al_x(\Phi_+^{\tout}(F))=\Phi_+^{\tout}(\al_x(F)),\quad \al_x(\Phi_-^{\tin}(F))=\Phi_-^{\tin}(\al_x(F))$ for $x\in\mcW$.

\end{enumerate}
\eel
%%%%%%%%%%%%%%%%%%%%%%%%%%%%%%%%%%%%%%%%%%%%%%%%%%%%%%%%%%%%%%%%%%%%%%%%%%%%%%%%%%%%%%%%%
\proof Let us consider the first limit in (\ref{minus-in}). 
Since there holds the estimate $\|F_{+}(h_T)\|\leq \|F\|\int dt\, |h(t)|$, it suffices to show the convergence
on the dense set of vectors $R'\Om$. First,  using the mean ergodic theorem, one checks that 
\beq
\underset{T\to\infty}{\slim}\textrm{ }F_{+}(h_T)\Om=P_+F\Om. \label{ergodic}
\eeq
In view of part (b) of the definition of the Borchers triple and the fact that $t_+\in\mcW$, we obtain that $F_{+}(h_T)\in\mcR$ for  $T$
sufficiently large. Hence, for any $F'\in R'$, 
\beq
\underset{T\to\infty}{\slim}\textrm{ }F_{+}(h_T)F'\Om=F'P_+F\Om,
\eeq
which proves the convergence. Since $\mcR$ is a von Neumann algebra, the limit $\Phi_+^{\tout}(F)$ is an element of $\mcR$.
Since $\Om$ is separating for $\mcR$,  this operator depends only on $\Phi_+^{\tout}(F)\Om=P_+F\Om$.

The second part of (\ref{minus-in}) is proven analogously. Property (a) follows by an application of the mean ergodic theorem, 
similarly as in (\ref{ergodic}). Property (b) is obvious from the definitions of $\Phi_+^{\tout}$, $\Phi_-^{\tin}$. \qed\\
%%%%%%%%%%%%%%%%%%%%%%%%%%%%%%%%%%%%%%%%%%%%%%%%%%%%%%%%%%%%%%%%%%%%%%%%%%%%%%%%%%%%%%%%%%%%%%%%%%%%%%%%%%%%
%%%%%%%%%%%%%%%%%%%%%%%%%%%%%%%%%%%%%%%%%%%%%%
%%%%%%%%%%%%%%%%%%%%%%%%%%%%%%%%%%%%%%%%%%%%%%%%%%%%%%%%%%%%%%%%%%%%%%%%%%%%%%%%%%%%%%%%%%%%%%%%%%%%%%%%%%%
Let us now define the asymptotic fields corresponding to the wedge $\mcW'$.
Keeping in mind that $J\mcR'J=\mcR$, we set for any  $F'\in\mcR'$ 
\beq
\Phi_+^{\tin}(F'):=J\Phi_+^{\tout}(JF'J)J, \quad \Phi_-^{\tout}(F'):=J\Phi_-^{\tin}(JF'J)J.
\eeq
Making use of formula~(\ref{inversion}), 
we easily obtain the following counterpart of  Lemma~\ref{right-wedge-lemma}.
%%%%%%%%%%%%%%%%%%%%%%%%%%%%%%%%%%%%%%%%%%%%%%%%%%%%%%%%%%%%%%%%%%%%%%%%%%%%%%%%%%%%%%%%%%%%%%%%%%%%%%%%%%
\bel \label{left-wedge-lemma} Let $F'\in \mcR'$.  Then 
\beqa
\Phi_+^{\tin}(F')=\underset{T\to-\infty}{\slim}\textrm{ }F'_{+}(h_T), \quad
\Phi_-^{\tout}(F')=\underset{T\to\infty}{\slim}\textrm{ }F'_{-}(h_T).\label{minus-in-in}
\eeqa
These operators depend only on the respective vectors $\Phi_+^{\tin}(F')\Om=P_+F'\Om$, $\Phi_-^{\tout}(F')\Om=P_-F'\Om$ and  satisfy
\begin{enumerate}
\item[(a)] $\Phi_+^{\tin}(F')\hil_+\subset\hil_+, \quad\Phi_-^{\tout}(F')\hil_-\subset\hil_-$,

\item[(b)] $\al_{x}(\Phi_+^{\tin}(F'))\!\!=\!\!\Phi_+^{\tin}(\al_{x}(F')),\quad \al_{x}(\Phi_-^{\tout}(F'))\!\!=\!\!\Phi_-^{\tout}(\al_{x}(F'))$ for $x\in\mcW'$.

\end{enumerate}

\eel
%%%%%%%%%%%%%%%%%%%%%%%%%%%%%%%%%%%%%%%%%%%%%%%%%%%%%%%%%%%%%%%%%%%%%%%%%%%%%%%%%%%%%%%%%%%%%%%%%%%%

Let us now proceed to the construction of scattering states.  
Clustering properties of the asymptotic fields are of importance here. Proceeding as in \cite{Bu75}, we note that for any $F,G\in\mcR$, $F',G'\in\mcR'$ there holds
\beqa
& & \!\!\!\!\!\!\!\!(\Phi_+^{\tout}(F)\Phi_-^{\tout}(F')\Om| \Phi_+^{\tout}(G)\Phi_-^{\tout}(G')\Om)\non\\
& &\ph{444444444444}=(\Phi_+^{\tout}(G)^*\Phi_+^{\tout}(F)\Om|\Phi_-^{\tout}(F')^*\Phi_-^{\tout}(G')\Om)\non\\
 & &\ph{444444444444}=(\Phi_+^{\tout}(F)\Om|\Phi_+^{\tout}(G)\Om)(\Phi_-^{\tout}(F')\Om|\Phi_-^{\tout}(G')\Om), \label{clustering-property}
\eeqa
where in the last step we made use of Lemma~\ref{right-wedge-lemma}~(a), Lemma~\ref{left-wedge-lemma}~(a) and of the fact that $\hil_+\cap [c\Om]^{\bot}$ is orthogonal to $\hil_-\cap[c\Om]^{\bot}$.
%%%%%%%%%%%%%%%%%%%%%%%%%%%%%%%%%%%%%%%%%%%%%%%%%%%%%%%%%%%%%%%%%%%%%%%%%%%%%%%%%%%%%%%%%%%%%%%%%%%%%%%%%%%%
Now for any $\Psi_+\in\hil_+$, (resp. $\Psi_-\in\hil_-$) we choose, with the help of property (c) of the Borchers triple, a sequence
$\{F_n\}_{n\in\nat}$ of elements of $\mcR$ (resp. a sequence $\{F_n'\}_{n\in\nat}$ of elements of $\mcR'$) s.t.
$\underset{n\to\infty}{\slim}\textrm{ }P_+F_n\Om=\Psi_+$ (resp.  $\underset{n\to\infty}{\slim}\textrm{ }P_+F_n'\Om=\Psi_-$).
By relation~(\ref{clustering-property}), the limit
\beq
\Psi_+\pout \Psi_-:=\underset{n\to\infty}{\slim}\textrm{ }\Phi_+^{\tout}(F_{n})\Phi_-^{\tout}(F'_{n})\Om \label{limiting}
\eeq 
exists and does not depend on the choice of the sequences within the above restrictions. We will call it
the outgoing scattering state. Next, we define the incoming scattering states as follows 
\beq
\Psi_+\pin\Psi_-:=J((J\Psi_+)\pout (J\Psi_-)). \label{incoming}
\eeq 
This definition is meaningful, since relation~(\ref{inversion}) gives $J\hil_+\subset\hil_+$ and $J\hil_-\subset \hil_-$. It is easily seen, that for suitable sequences $\{G_n\}_{n\in\nat}$ (resp. $\{G'_n\}_{n\in\nat}$) of elements of $\mcR$ (resp. of $\mcR'$), there holds
\beq
\Psi_+\pin\Psi_-=\underset{n\to\infty}{\slim}\textrm{ }\Phi_+^{\tin}(G'_{n})\Phi_-^{\tin}(G_{n})\Om, \label{in-limiting}
\eeq
similarly as in (\ref{limiting}).  The states constructed above have the following basic properties which justify their interpretation as scattering states:
%%%%%%%%%%%%%%%%%%%%%%%%%%%%%%%%%%%%%%%%%%%%%%%%%%%%%%%%%%%%%%%%%%%%%%%%%%%%%%%%%%%%%%%%%%%%%%%%%%%%%%%%%%%%%%
%%%%%%%%%%%%%%%%%%%%%%%%%%%%%%%%%%%%%%%%%%%%%%%%%%%%%%%%%%%%%%%%%%%%%%%%%%%%%%%%%%%%
\bel\label{lemma-three} For any $\Psi_\pm,\Psi_\pm'\in\hil_\pm$ 
\begin{enumerate}
\item[(a)] $(\Psi_+\pout \Psi_-,\Psi'_+\pout \Psi'_-)=(\Psi_+,\Psi'_+)(\Psi_-,\Psi'_-)$,
\item[(b)] $U(x)(\Psi_+\pout \Psi_-)=(U(x)\Psi_+)\pout (U(x)\Psi_-)$, for $x\in\real^2$.
\end{enumerate}
Analogous relations hold for  the incoming scattering states. 
\eel
%%%%%%%%%%%%%%%%%%%%%%%%%%%%%%%%%%%%%%%%%%%%%%%%%%%%%%%%%%%%%%%%%%%%%%%%%%%%%%%%%%%%%%
\proof Part (a) follows immediately from relation~(\ref{clustering-property}). As for part (b), for 
any $x\in\real^2$ we can choose such $y\in\mcW$ and $y'\in\mcW'$ that $x+y\in \mcW$ and $x+y'\in \mcW'$.   
We choose a sequence $\{F_n\}_{n\in\nat}$ of elements of $\mcR$ and $\{F_n'\}_{n\in\nat}$ of elements of $\mcR'$ s.t. $\underset{n\to\infty}{\slim}\textrm{ }P_+F_n(y)\Om=\Psi_+$ and $\underset{n\to\infty}{\slim}\textrm{ }P_-F_n'(y')\Om=\Psi_-$. Then
\beqa
U(x)(\Psi_+\pout \Psi_-)&=&\underset{n\to\infty}{\slim}\textrm{ }
\al_x(\Phi_+^{\tout}(F_{n}(y)))\al_x(\Phi_-^{\tout}(F'_{n}(y')))\Om\non\\
&=&\underset{n\to\infty}{\slim}\textrm{ }\al_{x+y}(\Phi_+^{\tout}(F_{n}))\al_{x+y'}(\Phi_-^{\tout}(F'_{n}))\Om\non\\
&=&\underset{n\to\infty}{\slim}\textrm{ }\Phi_+^{\tout}(F_{n}(x+y) )\Phi_-^{\tout}(F'_{n}(x+y'))\Om,
\eeqa
where we applied Lemma~\ref{right-wedge-lemma}~(b) and Lemma~\ref{left-wedge-lemma}~(b) in the second and third step. 
We note that the last state on the r.h.s. above
is just $(U(x)\Psi_+)\pout (U(x)\Psi_-)$, completing the proof of (b). 
The statement concerning the incoming states follows immediately from the properties of the outgoing states and from definition~(\ref{incoming}). \qed\\
%%%%%%%%%%%%%%%%%%%%%%%%%%%%%%%%%%%%%%%%%%%%%%%%%%%%%%%%%%%%%%%%%%%%%%%%%%%%%%%%%%%%%%%%%%%%%
After this preparation, we introduce the scattering subspaces
\beq
\hil^{\tin}=\hil_+\pin\hil_- \textrm{ and } \hil^{\tout}=\hil_+\pout \hil_-,
\eeq
which are spanned by the respective scattering states. In view of Lemma~\ref{lemma-three}, they
are canonically isomorphic to the tensor product $\hil_+\otimes\hil_-$. Similarly as in \cite{Bu75}, 
we define the scattering operator 
$S: \hil^{\tout}\to\hil^{\tin}$, extending by linearity the following
relation: 
\beq
S(\Psi_+\pout \Psi_-)=\Psi_+\pin\Psi_-. \label{scattering-matrix}
\eeq
By Lemma~\ref{lemma-three} this map is an isometry and it is invariant under translations. If
$S$ differs from (a constant multiple of) the identity transformation on $\hil^{\tin}$, then we say
that the theory is interacting. If $\hil^{\tin}=\hil^{\tout}=\hil$, then we say that the theory is asymptotically complete. 
In the next two sections we exhibit a class of theories which 
satisfy these two properties.

To conclude this section, we point out that the asymptotic fields form new Borchers triples,
which are non-interacting and asymptotically complete.
In view of Lemma~\ref{right-wedge-lemma}~(a) and Lemma~\ref{left-wedge-lemma}~(a),  we can define the following von Neumann algebras acting on $\hil^{\tas}:=\hil_{+}\otimes\hil_-$:
\beqa
\mcR^{\te{as}}&:=&\{\Phi_{+}^{\tout}(F)|_{\hil_{+}}\otimes \Phi_{-}^{\tin}(G)|_{\hil_{-}}  \,|\, F, G\in\mcR\, \}'',\label{first-asymptotic}\\ 
(\mcR')^{\te{as}}&:=&\{\Phi_+^{\tin}(F')|_{\hil_{+}}\otimes \Phi_-^{\tout}(G')|_{\hil_{-}}    \,|\, 
F', G'\in\mcR' \, \}''. \label{second-asymptotic}
\eeqa
Moreover, we set $U^{\te{as}}(x)=U(x)|_{\hil_+}\otimes U(x)|_{\hil_-}$ and $\Om^{\te{as}}=\Om\otimes\Om$.  
Clearly, there holds
\beq
\spec\, U^{\te{as}}=\spec\,U|_{\hil_+}+\spec\,U|_{\hil_-}\subset V_+.
\eeq
and $\Om^{\te{as}}$ is the  unique (up to a phase) unit vector which is invariant under the action of $U^{\tas}$.
Since $\Om^{\te{as}}$ is cyclic for  $\mcR^{\te{as}}$
and $(\mcR')^{\te{as}}$, and $(\mcR')^{\te{as}}\subset (\mcR^{\te{as}})'$, we obtain
that  $(\mcR^{\te{as}}, U^{\te{as}}, \Om^{\te{as}})$ 
is a Borchers triple w.r.t. $\mcW$. We call it the asymptotic Borchers triple of $(R,U,\Om)$. 
It has the following properties:
%%%%%%%%%%%%%%%%%%%%%%%%%%%%%%%%%%%%%%
\bep\label{asymptotic-BT} The Borchers triple $(\mcR^{\te{as}}, U^{\te{as}}, \Om^{\te{as}})$ defined above
gives rise to an asymptotically complete and non-interacting wedge-local quantum field theory.
Moreover, $\spec\,U^{\te{as}}=V_+$. 
\eep
%%%%%%%%%%%%%%%%%%%%%%%%%%%%%%%%%%%%%%%
\proof Making use of the fact that $(\mcR')^{\te{as}}\subset (\mcR^{\te{as}})'$, we obtain the equalities 
\beqa
& &\Phi_{+}^{\tout}(\Phi_{+}^{\tout}(F)|_{\hil_{+}}\otimes I )\Om^{\te{as}}=P_+F\Om\otimes\Om,\\
& &\Phi_{-}^{\tout}(I\otimes\Phi_{-}^{\tout}(F')|_{\hil_{-}} )\Om^{\te{as}}=\Om\otimes P_-F'\Om,
\eeqa
valid for any $F\in\mcR$, $F'\in\mcR'$. Thus we  conclude that $\hil^{\tas}_{+}\supset \hil_+\otimes [c\Om]$
and $\hil^{\tas}_{-}\supset  [c\Om]\otimes \hil_-$. Let $\Psi_{\pm}\in\hil_{\pm}$ and let
$\{F_n\}_{n\in\nat}$ (resp. $\{F_n'\}_{n\in\nat}$) be a sequence of elements of $\mcR$ (resp. $\mcR'$)
s.t.  $\underset{n\to\infty}\slim\, P_+F_n\Om=\Psi_+$ (resp. $\underset{n\to\infty}\slim\, P_-F_n'\Om=\Psi_-$). 
Then we get
\beqa
& &(\Psi_+\otimes\Om)\pout (\Om\otimes\Psi_-)\non\\
& &=\slim_n\Phi_{+}^{\tout}(\Phi_{+}^{\tout}(F_n)|_{\hil_{+}}\otimes I )
\Phi_{-}^{\tout}(I\otimes\Phi_{-}^{\tout}(F'_n)|_{\hil_{-}} )\Om^{\te{as}}\non\\
& &=\Psi_+\otimes\Psi_-. \label{asymptotic-outgoing}
\eeqa
By an analogous argument, we verify that 
\beq
(\Psi_+\otimes\Om)\pin (\Om\otimes\Psi_-)=\Psi_+\otimes\Psi_-. \label{asymptotic-incoming}
\eeq
We infer from equalities (\ref{asymptotic-outgoing}) and (\ref{asymptotic-incoming}) that $(\hil^{\tas})^{\tout}=(\hil^{\tas})^{\tin}=\hil^{\tas}$  (i.e. asymptotic completeness holds)
and $S=I$ (i.e. the theory is non-interacting). 

To justify the statement concerning the spectrum of $U^{\tas}$, we recall that $\hil_+\cap[c\Om]^{\bot}$ 
and $\hil_-\cap [c\Om]^{\bot}$ are assumed to be non-trivial. Consequently,  $\spec\, U|_{\hil_+}$ and
$\spec\, U|_{\hil_-}$ have some non-zero elements. From the existence of the unitary representation of
the Poincar\'e group $\ti U$, associated with the triple $(\mcR, U,\Om)$, we conclude that these
two spectra coincide with the right and left branch of the lightcone, respectively.  Since 
$\spec\, (U|_{\hil_+}\otimes U|_{\hil_-})=\spec\, U|_{\hil_+}+\spec\, U|_{\hil_-}$, the statement
follows. \qed

%%%%%%%%%%%%%%%%%%%%%%%%%%%%%%%%%%%%%%%%%%%%%%%

%%%%%%%%%%%%%%%%%%%%%%%%%%%%%%%%%%%%%%%%%%%%%%%%%%%%%%%%%%%%%%%%%%%%%%%%%%%%%%%%%%%%%%%%%
\section{Deformations and interaction}\label{deformation}
%%%%%%%%%%%%%%%%%%%%%%%%%%%%%%%%%%%%%%%%%%%%%%%%%%%%%%%%%%%%%%%%%%%%%%%%%%%%%%%%%%%%%%%%%

In the previous section we showed that for any Borchers triple in two-dimensional spacetime 
(with a unique vacuum state) we can canonically construct the scattering operator $S$ which
describes collisions of massless particles (or rather 'waves').   
In this section we consider a class of deformations of Borchers triples, introduced in \cite{BLS10},
and study their effect on the scattering operator. 
Similarly as in the massive case \cite{GL08,BS08}, the deformed theory turns out to be interacting, even if the original one is not. Moreover, we show that the property of asymptotic completeness is preserved under these deformations.

Let us recall briefly the deformation procedure of \cite{BLS10}. 
Let $(\mcR, U, \Om)$ be a Borchers triple w.r.t. the wedge $\mcW$. We denote by $\mcR^\infty$ the subset of
elements of $\mcR$ which are smooth under the action of $\al$ in the norm topology.
(It is easy to see that $\mcR^{\infty}$ is a  dense subalgebra of $\mcR$ in the strong operator topology). 
Let $\mcD$ be the dense domain of vectors which are smooth w.r.t. to
the action of $U$. Then, as shown in \cite{BLS10}, one can define for any $F\in\mcR^{\infty}$, 
and a matrix $Q$, antisymmetric w.r.t. the Minkowski scalar product $(x,y)\to xy$, the warped convolution
\beq
F_Q=\int dE(x)\,\al_{Qx}(F)
:=\lim_{\eps\searrow 0}(2\pi)^{-2}\int d^2x\, d^2y\, f(\eps x,\eps y) e^{-ixy}\al_{Qx}(F)U(y) \label{warped}
\eeq 
where $dE$ is the spectral measure of $U$ and $f\in S(\real^2\times\real^2)$ is s.t. $f(0,0)=1$.
The limit exists in the strong sense on vectors from $\mcD$ and is independent of the function $f$
within the above restrictions. We set
\beq
\mcR_{Q}:=\{\, F_Q\,|\, F\in\mcR^{\infty}\,\}''.
\eeq
Let us now restrict attention to the following family of  matrices 
\beqa
Q_{\ka}=\left(\begin{array}{cc} 0 & \ka \\
 \ka & 0 \end{array} \right),
\eeqa
where $\ka>0$, and recall a result from \cite{BLS10}:
%%%%%%%%%%%%%%%%%%%%%%%%%%%%%%%%%%%%%%%%%%%%%%%%%%%%%%%%%%%%%%%%%%%%%%%%%%%%%%
\bet\label{BLS-theorem} If $(\mcR, U, \Om)$ is a Borchers triple w.r.t. $\mcW$, then 
$(\mcR_{Q_{\ka}}, U, \Om)$ is also a Borchers triple w.r.t. $\mcW$. Moreover, 
$(\mcR')_{-{Q}_{\ka}}\subset (\mcR_{Q_{\ka}})'$.
\eet
%%%%%%%%%%%%%%%%%%%%%%%%%%%%%%%%%%%%%%%%%%%%%%%%%%%%%%%%%%%%%%%%%%%%%%%%%%%%%%%
Our goal is to  express the scattering operator
$S_{\ka}$ of the deformed theory $(\mcR_{Q_{\ka}}, U, \Om )$ by the scattering operator $S$ of the original
theory $(\mcR, U, \Om)$.  To this end, we prove the  following fact.
%%%%%%%%%%%%%%%%%%%%%%%%%%%%%%%%%%%%%%%%%%%%%%%%%%%%%%%%%%%%%%%%%%%%%%%%%%%%%%%%%%%%%%
 \bet\label{states-deformations} For any $\Psi_{\pm}\in\hil_{\pm}$  the following relations hold:
 \beqa
 & &\Psi_+\kpout \Psi_-=e^{-i\h\ka (H^2-P^2)}(\Psi_+\pout \Psi_-),\label{out}\\
 & &\Psi_+\kpin \Psi_-=e^{i\h\ka (H^2-P^2)}(\Psi_+\pin \Psi_-), \label{in}
 \eeqa
 where on the l.h.s. (resp. r.h.s.) there appear the scattering states of the deformed (resp. undeformed) theory. 
  \eet
 %%%%%%%%%%%%%%%%%%%%%%%%%%%%%%%%%%%%%%%%%%%%%%%%%%%%%%%%%%%%%%%%%%%%%%%%%%%%%%%%%%%%%%
\proof Let us first prove relation (\ref{out}).  
 To this end, we pick $F\in\mcR^\infty$,
 $F'\in(\mcR')^{\infty}$.  We set $\Psi_+=P_+F\Om=P_+F_{\Q}\Om$ and $\Psi_-=P_-F'\Om=P_-F'_{-\Q}\Om$,
 where we exploited the translational invariance of the state $\Om$.
 Since $F_{\Q}\in\mcR_{\Q}$ and, by Theorem~\ref{BLS-theorem}, $F'_{-\Q}\in(\mcR_{\Q})'$, the outgoing state of
 the deformed theory is given by
\begin{align}
\Psi_+\kpout \Psi_- 
 =\lim_{T\to\infty}F_{\Q,+}(h_T)F'_{-\Q,-}(h_T)\Om
 =\lim_{T\to\infty}F_{\Q,+}(h_T)F'_{-}(h_T)\Om\non\\
=\lim_{T\to\infty}\lim_{\eps\searrow 0}(2\pi)^{-2}\int d^2x\, d^2y\, f(\eps x,\eps y) e^{-ixy}\al_{Qx}(F_{+}(h_T))F'_-(h_T)(y)\Om, \label{deformed-outgoing}
\end{align}
where in the last step we made use of the fact that $F'_-(h_T)\Om\in\mcD$, and that 
$\Om$ is invariant under translations. To exchange the order of the limits, we use methods from the proof of Lemma 2.1 of \cite{BLS10}:
We note that for each polynomial $(x,y)\to L(x,y)$, there exists a polynomial $(x,y)\to K(x,y)$
s.t 
\beq
L(x,y)e^{-ixy}=K(-\pa_x,-\pa_y)e^{-ixy}.
\eeq
We choose $L$ so that $L^{-1}$ and its derivatives are absolutely integrable. Denoting temporarily  
$\Psi_T(x,y):=\al_{Qx}(F_{+}(h_T))F'_-(h_T)(y)\Om$,
we obtain 
\beqa
& &\ph{44}\lim_{\eps\searrow 0}(2\pi)^{-2}\int d^2x\, d^2y\, f(\eps x,\eps y) e^{-ixy} \Psi_T(x,y)  \non\\
& &=\lim_{\eps\searrow 0}(2\pi)^{-2}\int d^2x\, d^2y\,  e^{-ixy}K(\pa_x,\pa_y)f(\eps x,\eps y) L(x,y)^{-1}\Psi_T(x,y)\non\\
& &\ph{444444444}=(2\pi)^{-2}\int d^2x\, d^2y\,  e^{-ixy}K(\pa_x,\pa_y) L(x,y)^{-1}\Psi_T(x,y),
\label{mollifiers}
\eeqa
where in the first step we integrated by parts and in the second step we applied
the dominated convergence theorem. To obtain the last expression, we used the fact that
derivatives of $(x,y)\to f(\eps x, \eps y)$ contain powers of $\eps$ and thus vanish
in the limit. Substituting this expression to formula~(\ref{deformed-outgoing}) and making use again of
the dominated convergence theorem,  we arrive at
\beqa
\Psi_+\kpout \Psi_-\ph{44444444444444444444444444444444444444444444444444}\non\\
=(2\pi)^{-2}\int d^2x\, d^2y\,  e^{-ixy}K(\pa_x,\pa_y) L(x,y)^{-1} 
(U(Qx)\Psi_+)\pout (U(y)\Psi_-). \label{deformed-outgoing-state}
\eeqa
To interchange the limit $T\to\infty$ with the action of the derivatives, we exploited the fact that for any
$F_1\in\mcR^{\infty}$, $\mu\in\{0,1\}$, the derivative $\pa_{x^{\mu}}F_1:= (\pa_{x^{\mu}}F_1(x))|_{x=0}$ 
is an element of $\mcR^{\infty}$ and $\Phi_+^{\tout}(\pa_{x^{\mu}}F_1)(x)=\pa_{x^{\mu}}\Phi_+^{\tout}(F_1)(x)$. This  equality (as well as its counterpart for $\Phi_-^{\tout}$) follows immediately from the norm continuity of the respective map.

To analyze expression~(\ref{deformed-outgoing-state}), we will exploit some special features of massless theories
in two dimensions. First, we recall that $(H-P)\Psi_+=0$, and therefore
 \beqa
 U(\Q x)\Psi_+=e^{i\ka(Hx^1-Px^0)}\Psi_+
 =e^{-i\h\ka (H+P)(x^0-x^1)}\Psi_+.
 \eeqa
 Similarly, since $(H+P)\Psi_-=0$, we obtain
 \beq
 U(y)\Psi_-=e^{i\h(H-P)(y^0+y^1)}\Psi_-.
 \eeq
 Hence, exploiting the equalities $(H\pm P)\Psi_{\mp}=0$ and Lemma~\ref{lemma-three}~(b), we get
 \beqa
 (U(Qx)\Psi_+)\pout (U(y)\Psi_-)&=&e^{-\frac{i}{2}\ka (H+P)(x^0-x^1)}e^{\frac{i}{2}(H-P)(y^0+y^1)}(\Psi_+\pout\Psi_-)\non\\
 &=&U(v(x,y))(\Psi_+\pout\Psi_-),
 \eeqa
 where $v(x,y)=(\h(y^0+y^1-\ka x^0+\ka x^1), \h(y^0+y^1+\ka x^0-\ka x^1))$. 
 We substitute this expression to formula (\ref{deformed-outgoing-state}), obtaining
 \beqa
 (\Psi_+\kpout \Psi_-) \ph{4444444444444444444444444444444444444444444444}\non\\
=(2\pi)^{-2}\int d^2x\, d^2y\,  e^{-ixy}K(\pa_x,\pa_y) L(x,y)^{-1}
  U(v(x,y))  (\Psi_+\pout\Psi_-)\non\\
=\int\bigg(\lim_{\eps\searrow 0}(2\pi)^{-2}\int d^2x\, d^2y\,  e^{-ixy}f(\eps x,\eps y)
  e^{ip v(x,y)}\bigg) dE(p)(\Psi_+\pout\Psi_-). \label{long-formula}
 \eeqa
 Here in the second step we expressed $U(v(\cdot,\cdot))$ as a spectral integral
 and used the Fubini theorem to exchange the order of integration. 
 Then  we reversed the steps which led to formula~(\ref{mollifiers}). Now we
 analyze the function in the bracket above. Setting $p^{\pm}=\h(p^0\pm p^1)$, we get
 \begin{align}
  & (2\pi)^{-2}\int d^2x\, d^2y\,  e^{-ixy}f(\eps x,\eps y)e^{ip v(x,y)}\non\\
 =&(2\pi)^{-2}\int d^2x\, d^2y\,  
 f(\eps(x^0,x^1),\eps(y^0,y^1))
 e^{-i(\ka p^++y^0)x^0}e^{i(\ka p^++y^1)x^1}
 e^{ip^-(y^0+y^1)}\non\\
 =&(2\pi)^{-1}\int\, d^2y\, \eps^{-2}\hat f( -\eps^{-1} (\ka p^++y^0,\ka p^++y^1), 
 \eps(y^0,y^1))
 e^{ip^-(y^0+y^1)}\non\\
 =&(2\pi)^{-1}\int\, d^2y\, \hat f( -(y^0,y^1),\eps(\eps y^0-\ka p^+,\eps y^1-\ka p^+)) 
 e^{ip^-((y^0+y^1)\eps-2\ka p^+)}.\ph{4444}
 \end{align}
 Here $\hat f$ denotes the Fourier transform of $f$ w.r.t. the $x$ variable
 and in the last step we exploited the change of variables: 
 $(y^0,y^1)\to \big(\eps y^0-\ka p^+,\eps y^1-\ka p^+\big)$.
 Making use of the dominated convergence theorem, we  perform the limit
 $\eps\searrow 0$, obtaining
 \beq
 \lim_{\eps\searrow 0}(2\pi)^{-2}\int d^2x\, d^2y\,  e^{-ixy}f(\eps x,\eps y)e^{ip v(x,y)}
 =e^{-i\h\ka ((p^0)^2-(p^1)^2)}.
 \eeq
 In view of formula~(\ref{long-formula}), this completes the proof of (\ref{out}) for  dense sets of vectors $\Psi_{\pm}\in\hil_{\pm}$.
 For arbitrary $\Psi_{\pm}$ the statement follows by the limiting procedure~(\ref{limiting}). 
 
 The statement~(\ref{in}) concerning the incoming states can be shown using formula~(\ref{in-limiting}) and
an obvious modification of the above argument. We obtain it however more directly, using 
formula~(\ref{out}) and definition~(\ref{incoming}):
\beqa
\Psi_+\kpin \Psi_-=J((J\Psi_+)\kpout (J\Psi_-))&=&J\big(e^{-\frac{i}{2}\ka (H^2-P^2)}( (J\Psi_+)\pout (J\Psi_-) ) \big)\non\\
&=&e^{\frac{i}{2}\ka (H^2-P^2)}( \Psi_+\pin \Psi_-).
\eeqa   
Here in the last step we made use of the fact, shown in \cite{BLS10}, that the modular objects of the deformed and undeformed theory
coincide. We also exploited the relation $Jg(H,P)J= g(H,P)^*$, valid for any bounded, measurable function $g$, which follows from formula (\ref{inversion}). \qed\\
%%%%%%%%%%%%%%%%%%%%%%%%%%%%%%%%%%%%%%%%%%%%%%%%%%%%%%%%%%%%%%%%%%%%%%%%%%%%%%%%%%%%%%%%%%%%%%%%%%%%%%%%%%%%%%% 
%%%%%%%%%%%%%%%%%%%%%%%%%%%%%%%%%%%%%%%%%%%%%%%%%%%%%%%%%%%%%%%%%%%%%%%%%%%%%%%%%%%%%%%%%%%%%
We immediately obtain the following corollary:
%%%%%%%%%%%%%%%%%%%%%%%%%%%%%%%%%%%%%%%%%%%%%%%%%%%%%%%%%%%%%%%%%%%%%%%%%%%%%%%%
\bec\label{corollary} Let $S$  be the scattering operator of $(\mcR, U, \Om)$ and let $S_{\ka}$ be the scattering operator 
of $(\mcR_{Q_{\ka}}, U, \Om)$. Then
\beq
S_{\ka}=e^{i\ka (H^2-P^2)}S. \label{correction}
\eeq
In particular, if the original theory is asymptotically complete and non-interacting,
and $\spec\,U=V_+$,  then the deformed theory is asymptotically complete
and interacting.
%  Moreover $S_\ka$ has the crossing symmetry 
% for waves (cf. formula~(\ref{crossing-symmetry})), if $S$ does. 
\eec
%%%%%%%%%%%%%%%%%%%%%%%%%%%%%%%%%%%%%%%%%%%%%%%%%%%%%%%%%%%%%%%%%%%%%%%%%%%%%%%%%%%%%%%%%%%  
\proof Making use of Theorem~\ref{states-deformations}  and of the invariance of the scattering operator under translations, we obtain
\beqa
S_{\ka}(\Psi_+\kpout \Psi_-)&=&\Psi_+\kpin \Psi_-\non\\
&=&e^{i\h\ka (H^2-P^2)}(\Psi_+\pin \Psi_-)\non\\
&=&e^{i\h\ka (H^2-P^2)}S(\Psi_+\pout \Psi_-)\non\\
&=&e^{i\ka (H^2-P^2)}S(\Psi_+\kpout \Psi_-).
\eeqa
This proves formula~(\ref{correction}). The property of asymptotic completeness is preserved under the deformation,
since $e^{i\ka (H^2-P^2)}$ is a unitary. If $\hil^{\tout}=\hil$, $S=I$ and $\spec\,U=V_+$ then  $e^{i\ka (H^2-P^2)}$ is not a constant multiple of identity on $\hil^{\tout}$ i.e.  the deformed theory is interacting.  \qed\\
%%%%%%%%%%%%%%%%%%%%%%%%%%%%%%%%%%%%%%%%
%%%%%%%%%%%%%%%%%%%%%%%%%%%%%%%%%%%%%%%%%%%%%%%%%%%%%%%%%%%%%%%%%%%%%%%%%%%%%%%%%%%%%%%%%%%
%%%%%%%%%%%%%%%%%%%%%%%%%%%%%%%%%%%%%%%%%%%%%%%%%%%%%%%%%%%%%%%%%%%%%%%%%%%%%%%%%%%%%%%%%%%%%
%%%%%%%%%%%%%%%%%%%%%%%%%%%%%%%%%%%%%%%%%%%%%%%%%%%%%%%%%%%%%%%%%%%%%%%%%%%%%%%%%%%%%%%%%%%%%
We have shown in Proposition~\ref{asymptotic-BT} that any Borchers triple $(\mcR,U,\Om)$ with a unique vacuum vector $\Om$ and non-trivial single-particle subspaces $\hil_+\cap [c\Om]^{\bot}$, 
$\hil_-\cap [c\Om]^{\bot}$, gives
rise to an asymptotic Borchers triple $(\mcR^{\tas}, U^{\tas}, \Om^{\tas})$ which is asymptotically complete, non-interacting and s.t. $\spec\, U^{\tas}=V_+$. Hence, in view of the above corollary, the deformation of $(\mcR^{\tas}, U^{\tas}, \Om^{\tas})$ gives rise to an interacting, asymptotically complete theory.

 Interestingly, there exists a large class of Borchers triples which are unitarily equivalent to their asymptotic Borchers triples (in the sense of \cite{BLS10}). They give rise to interacting theories with a complete particle interpretation by a direct application of the deformation procedure. In the next section we show that the Borchers triples associated with  chiral conformal field theories belong to this class.

%%%%%%%%%%%%%%%%%%%%%%%%%%%%%%%%%%%%%%%%%%%%%%%%%%%%%%%%%%%%%%%%%%%%%%%%%%%

\section{Asymptotic completeness of chiral nets }\label{chiral}

%%%%%%%%%%%%%%%%%%%%%%%%%%%%%%%%%%%%%%%%%%%%%%%%%%%%%%%%%%%%%%%%%%%%%%%%%%%

In this section we consider a specific class of Borchers triples resulting from chiral nets.
We will show that such  triples are asymptotically complete, what is at first sight
surprising in view of the rich family of superselection sectors in chiral conformal field theory \cite{GF}. 
We recall, however, that in the present case particles (or rather 'waves') are composite objects, i.e. they 
may transform reducibly under the action of the Poincar\'e group.
Consequently, they may contain (pairs of) excitations from other sectors.

We start from the definition of a local net on $\real$, denoted by $(\ha\A, \ha U, \ha \Om)$.
It consists  of
\begin{itemize}
\item[(a)] a map $\real\supset \mcI\to\ha\A(\mcI)\subset B(\ha\hil)$, from open, bounded intervals
to von~Neumann algebras on  $\ha\hil$ s.t.
\beqa
& &\ha\A(\mcI)\subset \ha\A(\mcJ) \textrm{ for  } \mcI\subset \mcJ, \label{locality-condition} \\
& &[ \ha\A(\mcI),\ha\A(\mcJ)]=0 \textrm{ for } \mcI\cap \mcJ=\phi;
\eeqa 
  
\item[(b)] a unitary representation $\real\ni s\to \ha U(s)$ s.t. 
\beqa
& &\te{sp\,} \ha U\subset \real_+,\\
& &\ha U(s)\ha\A(\mcI) \ha U(s)^{-1}= \ha\A(\mcI+s)\, \textrm{ for } s\in \real;
\eeqa
\item[(c)] a unique (up to a phase)  unit vector $\ha\Om$, invariant under the action of $\ha U$, which is
cyclic  for any local algebra $\ha\A(\mcI)$. 

\end{itemize}

We remark, that there are many examples of local  nets on $\RR$. They arise, in particular, from conformal field theories on $S^1$ 
(see e.g. \cite{BMT, KL} for concrete examples). Given a theory on $S^1$ one obtains a net on the compactified real line by means of the  Cayley transform. Its restriction to the real line gives rise to a local net on $\RR$ with properties specified above.

Let $(\ha\A_1, \ha U_1, \ha\Om_1)$ and $(\ha\A_2,\ha U_2,\ha\Om_2)$ be two local nets on $\RR$,  and let $\ha\hil_1$, $\ha\hil_2$ be the respective Hilbert spaces.
We identify the two real lines with the lightlines $x+t=0$ and $x-t=0$
in $\RR^2$. To construct a local net  on $\RR^2$, acting on
the tensor product space $\H = \ha\H_1\otimes \ha\H_2$, we first specify the unitary representation of translations:
\beq
{U}(x):=\ha U_1\left(\frac{1}{\sqrt{2}}(x^0-x^1)\right)\otimes \ha U_2\left(\frac{1}{\sqrt{2}}(x^0+x^1)\right). 
\label{unitary-representation}
\eeq
Let $\al_{x}(\,\cdot\,):=U(x)\,\,\cdot\,\, U(x)^*$ be the corresponding group of translation 
automorphisms and let $\al^{(1/2)}_{x}(\,\cdot\, ):=\ha U_{1/2}(x)\,\,\cdot\,\,\ha U_{1/2}(x)^*$.  Then there holds
\beq
\alpha_{x}(A_1\otimes A_2) = \alpha^{(1)}_{\frac{1}{\sqrt{2}}(x^0-x^1)}(A_1)
\otimes \alpha^{(2)}_{\frac{1}{\sqrt{2}}(x^0+x^1)}(A_2), \quad A_1\in\ha\A_1,\, A_2\in\ha\A_2. \label{automorphisms}
\eeq
Any double cone in $\RR^2$ can be written as a product of two intervals on lightlines
$\mcI_1\times \mcI_2$. We define the corresponding local algebra  by $\B( \mcI_1\times \mcI_2) := \ha\A_1(\mcI_1)\otimes\ha\A_2(\mcI_2)$.
Setting $\Om=\hat\Om_1\otimes \hat\Om_2$, we obtain a triple $(\B, U,\Om)$, which we call a chiral net on $\RR^2$.  Defining
\beq
\mcR:=\bigvee_{ \mcI_1\times \mcI_2\subset \mcW}\B( \mcI_1\times \mcI_2), \label{mcR-def}
\eeq 
we arrive at a Borchers triple $(\mcR,U,\Om)$ associated with $(\B, U,\Om)$.

We will show that this Borchers triple is unitarily equivalent to its asymptotic Borchers triple $(\mcR^{\tas},U^{\tas},\Om^{\tas})$ and therefore, by Proposition~\ref{asymptotic-BT}, asymptotically complete and non-interacting. To this end, we determine the asymptotic fields in the following proposition:
%%%%%%%%%%%%%%%%%%%%%%%%%%%%%%%%%%%%%%%%%%%%%%%%%%%%%%%%%%%%%%%%%%%%%%%%%%%%%%
\bep\label{chiral-asymptotics} For any $A_1\in \ha\A_1(\mcI_1)$, $A_2\in\ha\A_2(\mcI_2)$ there holds
\beqa
\Phi_+^{\tout/\tin}(A_1\otimes A_2)&=&A_1\otimes (\ha\Om_2|A_2\ha\Om_2)I,\\
\Phi_-^{\tout/\tin}(A_1\otimes A_2)&=&(\ha\Om_1|A_1\ha\Om_1)I\otimes A_2.
\eeqa
(In the case of $\Phi_+^{\tout}$ and $\Phi_-^{\tin}$ it is assumed that $\mcI_1\times\mcI_2\subset\mcW$.
In the remaining cases $\mcI_1\times\mcI_2\subset\mcW'$).
\eep
%%%%%%%%%%%%%%%%%%%%%%%%%%%%%%%%%%%%%%%%%%%%%%%%%%%%%%%%%%%%%%%%%%%%%%%%%%%%%
\proof We consider only $\Phi_+^{\tout}$, as the remaining cases are analogous. From its definition and formula~(\ref{automorphisms}),  we obtain
\beq
\Phi_+^{\tout}(A_1\otimes A_2)=\underset{T\to\infty}{\slim}\textrm{ }A_1\otimes \int dt\, h_T(t)\al_{\sqrt{2}t}^{(2)}(A_2).
\eeq
We denote $A_2(h_T):=\int dt\, h_T(t)\al_{\sqrt{2}t}^{(2)}(A_2)$. This sequence has the following properties:
\beqa
\underset{T\to\infty}{\slim}\textrm{ }A_2(h_T)\ha\Om_2&=&(\ha\Om_2|A_2\ha\Om_2)\ha\Om_2, \label{auxiliary-one}\\
\lim_{T\to\infty}\|[A_2(h_T), A]\|&=&0, \textrm{ for any } A\in\ha\A_2(\mcI), \label{auxiliary-two}
\eeqa
where $\mcI$ is an arbitrary open, bounded interval. 
The first identity above follows from the mean ergodic theorem and the fact that $\ha\Om_2$ is the only vector invariant under the action of $\ha U_2$. The second equality is a consequence of the locality property~(\ref{locality-condition}). Now since $[\ha\A_2(\mcI)\ha\Om_2]=\ha\hil_2$, 
we obtain from relations~(\ref{auxiliary-one}), (\ref{auxiliary-two})
\beq
\underset{T\to\infty}{\slim}\textrm{ }A_2(h_T)=(\ha\Om_2|A_2\ha\Om_2)I.
\eeq
This completes the proof. \qed\\
%%%%%%%%%%%%%%%%%%%%%%%%%%%%%%%%%%%%%%%%%%%%%%%%%%%%%%%%%%%%%%%%%%%%%%%%%%%%%%%%%%%%%%%
Now we can easily prove the main result of this section:
%%%%%%%%%%%%%%%%%%%%%%%%%%%%%%%%%%%%%%%%%%%%%%%%%
\bet Any Borchers triple $(\mcR,U,\Om)$ associated with a chiral net on $\RR^2$
is unitarily equivalent to its asymptotic Borchers triple $(\mcR^{\tas},U^{\tas},\Om^{\tas})$.
More precisely, there exists a unitary map $W:\hil^{\tas}\to \hil$ s.t. $W \mcR^{\tas}=\mcR W$,
$WU^{\tas}(x)=U(x)W$ and $W\Om^{\tas}=\Om$.
\eet
%%%%%%%%%%%%%%%%%%%%%%%%%%%%%%%%%%%%%%%%%%%%%%%%%%
\proof  By cyclicity of  $\Om$ under the action of $\mcR$ and the mean ergodic
theorem (cf. formula~(\ref{ergodic})), there holds
\beq
\hil_{\pm}=[\,\Phi_{\pm}^{\tout}(F)\Om\,|\, F\in \mcR\,].
\eeq
Thus, applying  Proposition~\ref{chiral-asymptotics}, and exploiting the cyclicity of $\ha\Om_{1}$, $\ha\Om_2$ under
the action of the respective local algebras, we obtain  
\beqa
& &\hil_+=\ha\hil_1\otimes [c\ha\Om_2], \\
& &\hil_-=[c\ha\Om_1]\otimes \ha\hil_2. 
\eeqa
Recalling that $\hil^{\tas}=\hil_+\otimes\hil_-$ and $\hil=\ha\hil_1\otimes\ha\hil_2$, 
 we define a unitary map $W: \hil^{\tas}\to \hil$, extending by linearity the relation
\beqa
W((\Psi_1\otimes\ha\Om_2)\otimes (\ha\Om_1\otimes\Psi_2))=\Psi_1\otimes\Psi_2, \quad \Psi_1\in\ha\hil_1, \Psi_2\in\ha\hil_2.
\eeqa  
It is readily verified  that
\beqa
& &WU^{\tas}(x)=U(x)W, \\
& &W\Om^{\tas}=\Om,\\
& &W\{\Phi_+^{\tout}(A_1\otimes A_2)|_{\hil_+}\otimes\Phi_-^{\tin}(B_1\otimes B_2)|_{\hil_-}\}\non\\
& &\ph{4444444444444444}=     (\ha\Om_1|B_1\ha\Om_1) (\ha\Om_2|A_2\ha\Om_2) \{A_1\otimes B_2\}W,    \label{W}
\eeqa
where  $A_1\otimes A_2$, $B_1\otimes B_2$ comply with the assumptions of
Proposition~\ref{chiral-asymptotics}.   By definition~(\ref{mcR-def}), the
elements in the curly bracket on the r.h.s. of (\ref{W}) generate $\mcR$. 
Making use of this fact and of the identities 
 $\Phi_{+}^{\tout}(F)|_{\hil_{+}}=P_{+}F|_{\hil_{+}}$, 
 $\Phi_{-}^{\tin}(F)|_{\hil_{-}}=P_{-}F|_{\hil_{-}}$, where  $F\in\mcR$,
we obtain that the double commutant of the set of elements in the curly bracket
on the l.h.s. of  (\ref{W}) coincides with $\mcR^{\tas}$. Hence  $W\mcR^{\tas} =\mcR W$, which concludes the proof. \qed\\
%%%%%%%%%%%%%%%%%%%%%%%%%%%%%%%%%%%%%%%%%%%%%%%%%%%%%  
In view of the above theorem, we obtain from Proposition~\ref{asymptotic-BT}:
%%%%%%%%%%%%%%%%%%%%%%%%%%%%%%%%%%%%%%%%%%%%%%%%%%%%%
\bec Any Borchers triple $(\mcR, U,\Om)$, associated with a chiral net, gives rise to an asymptotically complete, non-interacting theory.  Moreover, $\spec\, U=V_+$.
\eec
%%%%%%%%%%%%%%%%%%%%%%%%%%%%%%%%%%%%%%%%%%%%%%%%%%%%%%
Hence, by Corollary~\ref{corollary}, deformations of such Borchers triples give rise to asymptotically complete, interacting theories.
%%%%%%%%%%%%%%%%%%%%%%%%%%%%%%%%%%%%%%%%%%%%%%%%%%%%%%%%%%%%%%%%%%%%%%%%%%%%%%%%%%%%%%%
%%%%%%%%%%%%%%%%%%%%%%%%%%%%%%%%%%%%%%%%%%%%%%%%%%%%%%%%%%%%%%%%%%%%%%%%%

\section{Concluding remarks}\label{conclusions}

%%%%%%%%%%%%%%%%%%%%%%%%%%%%%%%%%%%%%%%%%%%%%%%%%%%%%%%%%%%%%%%%%%%%%%%%%

In this paper we applied the deformation method, developed in \cite{BLS10}, 
to two-dimensional  massless theories. We have shown that the deformation procedure
not only introduces interaction, as expected from the massive case \cite{GL08, BS08}, 
 but also preserves the property of asymptotic completeness. By deforming chiral conformal field theories, 
 we obtained a  large class of wedge-local theories, which are interacting and asymptotically complete.    
As the resulting scattering matrices  are Lorentz invariant, one can hope for the
existence of local observables in these models. We recall that  negative results, concerning this issue, have so far been  established only in spacetimes of  dimension larger than two \cite{BLS10}.

A large part of our investigation was devoted to scattering of massless particles ('waves') in two-dimensional
wedge-local theories. 
It turned out that the scattering theory developed in \cite{Bu75} for
local nets of observables generalizes naturally to the wedge-local framework: To construct the two-body scattering matrix, it suffices to know the Borchers triple. It is an interesting open problem, if this fact remains true for scattering of massless particles 
in spacetimes of higher dimension. We recall that for local nets of observables  scattering theory of massless excitations 
is well understood in the physical spacetime \cite{Bu77}.

\vspace{1cm}

\noindent\bf Acknowledgements. \rm
The authors would like to thank Prof.$\!$ D.$\!$ Buchholz, Dr.$\!$ G.$\!$ Lechner and
D.$\!$ Cadamuro for interesting discussions. Moreover, W.D. gratefully acknowledges 
 hospitality of the Erwin Schr\"odinger International Institute for Mathematical Physics in Vienna during the
final stages of this work.

%%%%%%%%%%%%%%%%%%%%%%%%%%%%%%%%%%%%%%%%%%%%%%%%%%%%%%%%%%%%%%%%%%%%%%%%%%%%%%%%%%%%%%%%

\end{document}